# Superconductivity in palladium hydride and deuteride at 52–61 kelvin

H. M. Syed, T. J. Gould, C. J. Webb and E. MacA. Gray*

Queensland Micro- and Nanotechnology Centre, Griffith University, Nathan 4111, Brisbane, Australia

*e-mail: e.gray@griffith.edu.au

**Abstract**

We report the observation of conventional superconductivity at the highest temperature yet attained without mechanical compression, around 54 kelvin in palladium-hydride and 60 kelvin in palladium-deuteride. The remarkable increase in $T_c$ compared to the previously known value was achieved by rapidly cooling the hydride and deuteride after loading with hydrogen or deuterium at elevated temperature. Our results encourage hope that conventional superconductivity under ambient conditions will be discovered in materials with very high hydrogen density, as predicted more than a decade ago.

**1. Introduction**

Conventional superconductivity at about ten kelvin was discovered in palladium-hydride and palladium-deuteride in 1972[1,2], raising hopes that the high density of hydrogen in metal-hydrides and compounds containing a high density of hydrogen might lead to metallic hydrogen behaviour[3]. No high-temperature hydride superconductors were found. Prior to the 1986 discovery of high-temperature superconductivity in cuprate ceramics[4], for which the current record superconducting transition temperature ($T_c$) is 164 K under high mechanical pressure[5], the highest recorded $T_c$ was 22.4 K[6]. Attention then focused on non-conventional cuprates until the discovery of conventional superconductivity at 39 kelvin in magnesium-diboride[7], then on the 2008 discovery of iron-based superconductors[8], with $T_c$ values still not exceeding 58 K despite intense research effort[9]. Recently Drozdov et al.[10] found that 140 GPa mechanical pressure induced conventional superconductivity in sulfur-hydride at 203 K.

$PdH_x$ is superconducting[1] for $0.7 < x < 1.0$, with $T_c = 8-10\,\text{K}$ for $x \approx 1.0$ confirmed in many experiments[11]. Palladium deuteride, $PdD_x$, exhibits a reverse isotope effect[2], with $T_c = 10-12$ K for $x \approx 1.0$ confirmed in many experiments[11]. This record for an interstitial metal-hydride was not surpassed until now[11], although there have been unsubstantiated and disputed claims of superconductivity in PdH near room temperature[12]. The superconducting properties of most simple metals and some alloys can be qualitatively understood within the 1957 BCS (Bardeen–Cooper–Schrieffer) picture[13], in which conduction electrons interact via phonons to form correlated Cooper pairs. McMillan[14] extended the BSC picture for strongly coupled superconductors using Eliashberg's formulation and showed that the electron–phonon coupling increased with the strength of the electron–phonon interaction and the density of electronic states at the Fermi level, and also increased for decreasing ion mass and average phonon frequency. The latter dependency is counteracted by the direct proportionality of $T_c$ to the average phonon frequency. Historically, theoretical opinion wavered between an upper bound of 30–40 kelvin for phonon-mediated superconductivity[14,15] and no in-principle limit[3,16-18].

McMillan's theory was further developed by Allen and Dynes[17], to yield a more widely applicable formula for $T_c$. Contrary to McMillan, Allen and Dynes found no in-principle upper bound to $T_c$. They concluded that to increase $\lambda$ it is more important to increase the strength of the electron–phonon interaction than to decrease the average phonon frequency. Further developments led to a formalism that has had reasonable success in calculating values for $T_c$ from first principles for conventional superconductors[19-24].

The superconducting properties of $PdH_x$/$PdD_x$ at about 10 K can be understood within the Eliashberg formulation of the BCS picture, despite the counterintuitive reverse isotope effect, which is a consequence of anharmonicity of the interatomic potential seen by H/D[21].

Pd metal readily absorbs hydrogen at room temperature and less than 0.1 MPa gas pressure. H/D occupies octahedral (*oct*) interstitial sites in the face-centred cubic Pd lattice. At first a solid solution, the α phase, is formed. As the hydrogen pressure increases, the concentrated β phase, $PdH_{0.6}$ is formed. The phase transformation is accompanied by lattice expansion and decrease of the magnetic susceptibility as the hole in the Pd $d$ band is filled. The lattice mismatch between the α and β phases causes the generation of dislocations[25].

Achieving $x \approx 1.0$ in $PdH_x/D_x$ at room temperature requires the application of extreme hydrogen pressures (hundreds of MPa) or electrolytic charging[11]. For experiments on superconductivity at low temperatures, $PdH_x/D_x$ typically has been prepared by electrolysing Pd at room temperature and cooling rapidly to cryogenic temperatures to minimise H/D loss following removal from the electrolysis cell[11]. This procedure involves passage through the two-phase region and so leaves the sample with a high concentration of dislocations. The role, if any, of dislocations in superconductivity in $PdH_x/D_x$ is not clear, although it has been proposed[26] that filamentary superconductivity may occur within dislocation cores.

The origin of the reverse isotope effect ($T_c(D) > T_c(H)$) in PdH/D at low temperatures can be understood in the following way. The potential at the *oct* site is strongly anharmonic, with a positive anharmonicity parameter, meaning that the interstitial atom sees a hardening potential as its total energy increases. The effect is to shift the bulk of the pDOS associated with H/D to higher frequencies. As shown by Errea et al.[27], this increases $\omega_{log}$ (see §2.2.) somewhat, but also strongly decreases $\lambda$, with the result that $T_c$ is suppressed by a factor approaching ten for H. The heavier D interstitial sits lower in its potential well and sees a somewhat softer potential, so that anharmonicity reduces $T_c$ by a smaller factor than for H, resulting in a positive isotope effect.

Here we report the observation of superconductivity at around 54 kelvin in palladium-hydride and 60 kelvin in palladium-deuteride. The remarkable increase in $T_c$ compared to the previously known value was achieved by rapidly cooling the hydride and deuteride after loading with hydrogen or deuterium at elevated temperature. Our results encourage hope that conventional superconductivity under ambient conditions will be discovered in materials with very high hydrogen density, as predicted more than a decade ago[28].

## 2. Methods

### 2.1. Experiments

Superconductivity was detected by measuring the resistance of the sample. The experiments were conducted with the sample contained in a cylindrical pressure cell made from 316 stainless steel tube with a thinned lower section machined from 660 stainless steel. The higher tensile strength of 660 compared to 316 helped to minimise the thermal mass of the sample environment. The pressure rating was 10 MPa at 300°C. The cell was 12.7 mm in diameter × 1200 mm long, so that it could be inserted in a furnace or cryostat. The cell was equipped with feed-throughs for measuring the sample resistance and temperature.

The sample was connected by a flexible pipe to a gas manifold equipped with metal-hydride sources of hydrogen and deuterium and a pressure transducer. This connection was maintained throughout the experiment, which allowed the pressure over the sample to be maintained during cooling by admitting additional gas. A small tubular furnace was used to heat the sample cell above room temperature. A Cryogenics Limited cryostat cooled by a two-stage closed-cycle refrigerator was used to cool the sample cell below room temperature. The base sample temperature reached with the pressure cell inserted was approx. 40 K.

The sample was Pd wire with diameter 0.5 mm and length approx. 80 mm, formed into a serpentine shape on an alumina substrate small enough to fit into the 10-mm bore of the pressure cell. Copper leads were spot-welded to the sample. The sample temperature was measured with a high-quality Pt thermometer (Lakeshore Cryotronics Pt102) mounted in close proximity to the sample. The temperature value was calculated by using the DIN IEC 751 Temperature/Resistance table for platinum sensors. Both resistors were connected in four-probe configuration to minimise the effect of lead resistance. Once the sample had been assembled onto its substrate and the leads attached, the cell was sealed and all subsequent treatments and measurements took place in situ, so that the sample was exposed to only vacuum or $H_2/D_2$ thereafter.

The sample was initially annealed in situ at 500°C under vacuum for 8 hours. It was then heated to approximately 300°C and loaded with H/D from the gas phase, typically at 10 MPa pressure. The sample cell was then extracted from the furnace and the lower portion was pre-cooled in liquid nitrogen. The pressure cell was then transferred to the cryostat and cooled as fast as possible to about 40 K. At 40 K, the estimated equilibrium H/D content was $x \approx 0.95$ [29]. It should be noted, however, that our measurements were non-equilibrium and the state of the H/D in Pd was unusual. The resistivity was measured during rapid heating to room temperature, taking approximately 30 minutes. Both rapid cooling and rapid heating were necessary to observe the superconducting transition. To unload H/D from the sample, it was re-heated to 300°C under pressure, depressurised and evacuated, once again avoiding the two-phase region.

Figure 1 shows the same experimental methodology applied to copper wire, to demonstrate the correct working of the apparatus and the absence of experimental artefacts, and to calibrate the sample temperature.

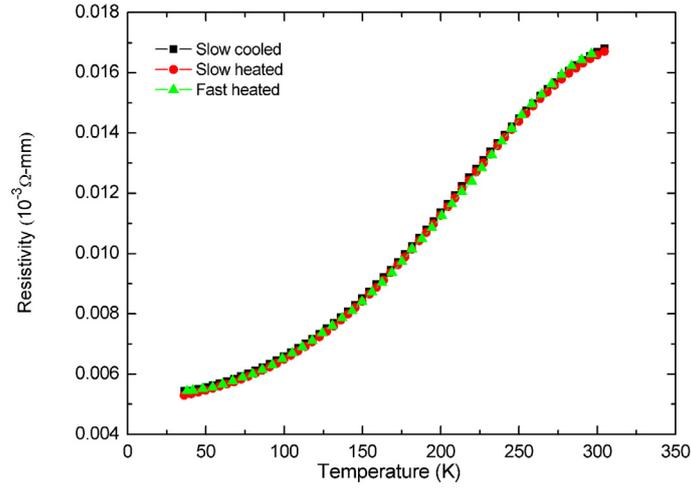

**Figure 1.** Commercial copper wire pressurised with 10 MPa hydrogen at 300°C, cooled to 40 K under 10 MPa and heated quickly or slowly as indicated.

*2.2. Theory*

In McMillan's modified BCS picture for strongly coupled superconductors[14], the electron–phonon coupling constant is

$$\lambda = \frac{N(0)\langle I^2 \rangle}{M\langle \omega^2 \rangle} = \frac{\eta}{M\langle \omega^2 \rangle} \quad (1)$$

where $\lambda$ is the electron–phonon coupling constant, $N(0)$ is the density of electronic states at the Fermi level, $\langle I^2 \rangle$ is the mean-square matrix element of the electron–phonon interaction, $M$ is the ionic mass and $\langle \omega^2 \rangle$ is the (weighted) mean-square phonon frequency. Here $\lambda$ plays the same role as $N(0)V$ in the BCS picture and is defined in terms of the Eliashberg[14,30,31] spectral function $\alpha^2(\omega)F(\omega)$:

$$\lambda = 2\int_0^\infty \frac{1}{\omega}\alpha^2(\omega)F(\omega)d\omega \quad (2)$$

Eqn (1) has been the basis for connecting theory and experiment. Its importance lies in the separation of the purely electronic contributions ($\eta$) in the numerator from the phonon contributions in the denominator.

McMillan's theory was further developed by Allen and Dynes[17], to yield a more widely applicable formula for $T_c$:

$$T_c \approx \frac{\omega_{log}}{1.2} \exp\left[-\frac{1.04(1+\lambda)}{\lambda - \mu^*(1+0.62\lambda)}\right] \quad (3)$$

where $\mu^*$ is the Coulomb coupling constant (pseudopotential). The logarithmically averaged phonon frequency is given by

$$\ln(\omega_{log}) = \frac{\int_0^\infty \frac{\ln(\omega)}{\omega} \alpha^2(\omega) F(\omega) d\omega}{\int_0^\infty \frac{1}{\omega} \alpha^2(\omega) F(\omega) d\omega} \quad (4)$$

Realistic *ab intito* calculations of $\alpha(\omega)$, which necessarily involve large supercells and a full accounting of difficult-to-converge phonon and higher order effects, are now possible and have recently been carried out for the low-temperature *oct* superconducting state[27]. To do so requires knowledge of the H/D configuration, which is not available for this new state with much higher $T_c$. The alternative approach of quantitatively analysing the experimental data above $T_c$ to estimate $\lambda$ is not feasible because of the variability of the resistivity within and between experimental trials. However, we *can* test the phonon density of states (pDOS) as a proxy for $\alpha(\omega)$ in reasonably sized 2×2 supercells for postulated H/D configurations.

All calculations were carried out in VASP[32-35] using 16×16×16 k-points and an energy cutoff of 700 eV. Phonons were calculated using 24×24×24 k-points and an energy cutoff of 800 eV. We used the PAW pseudopotentials for Pd and H and the PBE gradient approximation[36] for energies. Phonon band and density-of-states (pDOS) data was generated by Phonopy[37] using analytic forces from VASP, and were found to be well converged. The pDOS results were post-processed and plotted using customised python code.

## 3. Results and Discussion

### *3.1. Experiments*

Figure 2 compares the resistivites of $PdH_x/D_x$ cooled and heated quickly, demonstrating the existence of a zero-resistance state (within resolution) below about 54 K for the hydride and 60 K for the deuteride. Also shown are pure Pd and slow-cooled $PdH_x/D_x$, which do not exhibit any sudden change in resistivity. Also shown is $PdD_x$ cooled in a magnetic field, showing that the transition has been switched off. The highest measured $T_c$ was 55.6 K in $PdH_x$ and 61.4 K in $PdD_x$.

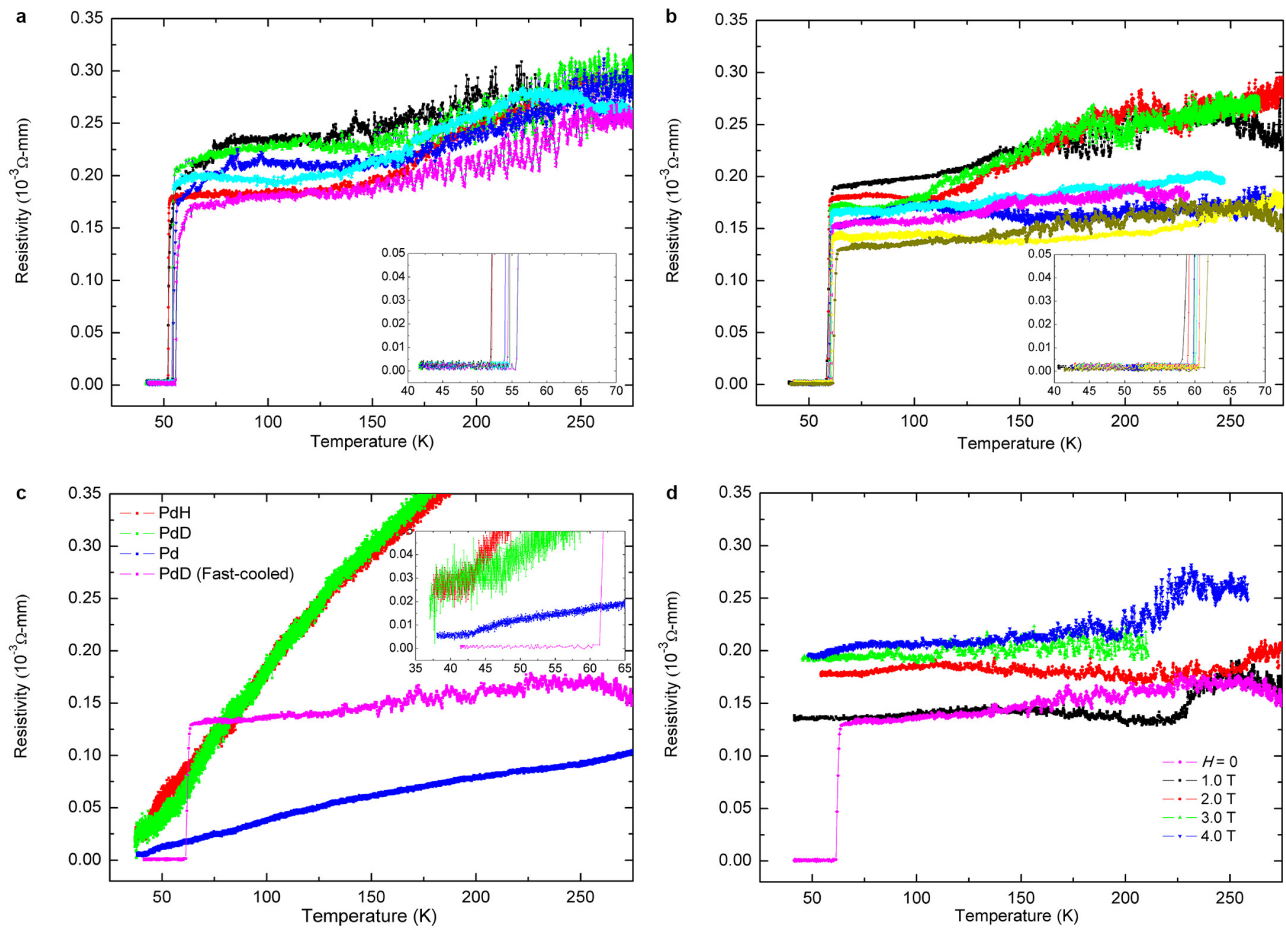

**Figure 2.** Resistivities of wire samples loaded with hydrogen/deuterium 300°C and cooled to approx. 40 K before measurement while heating to 300 K.

**(a)** $PdH_x$ rapidly cooled and rapidly heated. Superconductivity was observed below about 54 K.

**(b)** $PdD_x$ rapidly cooled and rapidly heated. Superconductivity was observed below about 60 K. Note the higher transition temperature compared to the hydride.

**(c)** $PdH_x$, $PdD_x$ and pure Pd, all cooled slowly and heated slowly, compared to rapidly cooled and rapidly heated $PdD_x$. Note in the latter case the transition to a state with lower resistivity than the pure metal. Note also the smooth variation of resistivity with temperature and the high contribution of H/D to the resistivity.

**(d)** $PdD_x$ rapidly cooled and rapidly heated in a magnetic field as indicated. Note the quenching of the superconducting transition by the magnetic field.

The observation of superconductivity in palladium hydride/deuteride at some five times the temperature previously known is highly remarkable.

Four pieces of evidence point convincingly to the observed transition being real and not an artefact of the experimental apparatus or methodology: (i) Measurements on copper wire treated exactly the same as the Pd wire (Fig. 1) showed a very smooth variation with temperature and no anomaly. The resistivity measured during fast heating differed minimally from that measured during slow heating, confirming that only a small temperature difference develops between the sample and the thermometer under dynamic conditions. These measurements proved that the apparatus worked very well. (ii) Whenever the transition was observed, the resistance was zero within resolution at lower temperatures. This could only be because either (a) a sufficient volume fraction of the sample was superconducting to give overall zero resistance, or (b) a short-circuit sometimes occurred between the voltage-sensing wires somewhere, always in the same temperature range. (iii) Repeating the experiments with deuterium instead of hydrogen led to the same transition being observed, but always at a higher temperature (Fig. 2 (b)). It is unreasonable that this isotope effect could be an artefact of the equipment. Therefore possibility (b) above should be discarded. (iv) The zero-resistance state was never observed when the experiments were repeated with a magnetic field of 1 T or higher applied to the sample (Fig. 2(d)). It is unreasonable that this effect could be an artefact of the experiment owing to possibility (b) above.

A major question is why high-temperature superconductivity was never observed when the sample was cooled and heated slowly. Cooling slowly is well known to produce a low-temperature superconducting state at around 10 K for a hydrogen-to-metal ratio approaching 1, surveyed in detail in [11]. Low-temperature superconductivity in this system is associated with occupancy of octahedral interstitial sites. This is known from neutron diffraction measurements with the deuteride (summarised in [38]), which show convincingly that D (and presumably H) occupy octahedral interstices in the Pd lattice at and below room temperature.

Loss of hydrogen from the sample has been given as a reason for variable $T_c$ reported for $PdH_x/D_x$ in the literature, but in our experiments the $H_2/D_2$ pressure was always maintained high enough during cooling for the sample to remain in the pure β phase, and in fact to gain hydrogen as it cooled, since the equilibrium pressure for a given concentration of H/D in a metal falls very rapidly with decreasing temperature. Less hydrogen would diffuse out of the sample during rapid heating compared to slow heating, but less hydrogen would have diffused into the sample anyway during rapid cooling compared to slow cooling. While the transition to low-temperature superconductivity at about 10 K is very sensitive to the actual H/D concentration in the sample[11], the transition to high-temperature superconductivity seems to be either observed or not observed, with a quite small relative variation in the value of $T_c$. Variable H/D concentration probably accounts for the variation in the measured $T_c$ (several K in 50-odd K over many trials), but prevention of hydrogen loss from the sample is not the mechanism by which rapid cooling followed by rapid heating causes (or allows) the transition to appear.

We conclude that rapidly cooling the hydrogenated sample from high temperature freezes-in a high-temperature hydrogen configuration that is conducive to superconductivity at 52–61 K, and rapid heating preserves this state while the measurement takes place. The nature of this state needs to be determined. As we argue below, experimental and theoretical evidence suggests that in this high-temperature state some H/D atoms occupy tetrahedral (*tet*) instead of, or in addition to, octahedral (*oct*) interstices.

It is known from neutron diffraction measurements with the deuteride that ordering of the octahedral D occurs at temperatures below about 100 K[39], down to about 50 K, although on time scales much longer than that of the present experiments. The influence of D ordering – or its absence – on the low-temperature superconductivity is not known, but it almost certainly plays no role in our observations of high-temperature superconductivity.

The variability in the transition temperature and resistivity above $T_c$ when the sample was cooled and heated quickly suggests that the configuration of hydrogen at about 40 K is very sensitive to the actual conditions and cooling rate. From the neutron diffraction results mentioned it is known that quenching the sample prevents D ordering, and so the possibility that the higher value of $T_c$ comes about because of the absence of D ordering needs to be considered. Rapid heating would avoid ordering while the sample remained below 100 K. It would be surprising, however, if higher values of $T_c$ were not observed in previous work, since cooling times from room temperature to 40 K and below comparable to, and even shorter than, the cooling time in the present experiments would have been possible. In this regard it should be noted that the cooling power of the closed-cycle refrigerator used for this work is much lower than the cooling power of a conventional liquid-He cryostat.

The possibility of H/D diffusion at cryogenic temperatures also needs to be considered in relation to the observed effect of the differing isotopes, with D-loaded Pd always exhibiting a higher $T_c$. If the quenched-in state is superconducting, then the slower diffusion of D relative to H could plausibly delay the return to the normal state until a higher temperature is reached, just as it delays the return to the equilibrium, non-superconducting state. Again noting the relatively small variation in the measured $T_c$ between successful trials with $PdH_x$, despite the wild variation in the resistivity above $T_c$, it seems unlikely that the consistently higher value of $T_c$ obtained for $PdD_x$ is entirely due to the slower diffusion of D. While this explanation cannot be definitively excluded without further experiments, taking the probable isotope effect together with the metallic character of the material, for the purpose of discussing and explaining the experimental results, we suppose that the superconductivity is phonon-mediated, there being no evidence to the contrary.

The wild variation of the resistivity at high temperatures requires explanation. Comparing measurements on pure Pd wire in vacuum (Fig. 2(c)) and on commercial Cu wire treated identically to the Pd samples including exposure to $H_2/D_2$ (Fig. 1) shows that this variability is real and not an experimental artefact. The extra electron scattering caused by interstitial H/D has a very big effect on the resistivity (Fig. 2(c)), so this observation suggests an inhomogeneous state in which the local hydrogen configuration fluctuates.

Aside from imperfections in the electrical connections to the sample, the resistance of a macroscopic sample will be zero if there is a continuous network of superconducting material between the two ends, even if the majority of the sample remains normal. The somewhat random occurrence or non-occurrence of a transition to zero resistivity suggests that the superconducting network is either continuous or broken, as in the percolation phenomenon in which the gross conductivity of a three-dimensional network of conducting elements is suddenly lost at a critical fraction of broken connections[40]. In three dimensions, the percolation threshold is very roughly 30%. We can then understand unsuccessful experiments as those in which, owing to the numerous variable experimental factors, the percolation threshold was not reached.

These ideas lead to an explanation of why was superconductivity was not always observed when the sample was cooled and heated quickly. The success rate with deuterium loading was much higher than with hydrogen. In unsuccessful trials, i.e. those in which the superconducting transition was not observed, once again, the resistivity varied wildly with temperature, and between trials (Fig. 3). Once again, this could indicate an inhomogeneous state, but one in which the sample as a whole does not reach a state of zero resistance. A further point is that in all cases where the sample was cooled and heated quickly (Fig. 2(a), (b), (d); Fig. 3), the resistivity at room temperature was no more than about half its equilibrium value (Fig. 2(c)), whether or not the superconducting transition was observed. This shows that the sample was always very far from equilibrium, even at room temperature, after rapid cooling and heating, supporting the idea that a non-equilibrium state has been frozen-in by rapid cooling from 300 °C and is prevented from transforming to the equilibrium state by the limited diffusion rates of H and D at cryogenic temperatures, with the slower diffusion of D allowing percolation of the gross superconducting state more reliably.

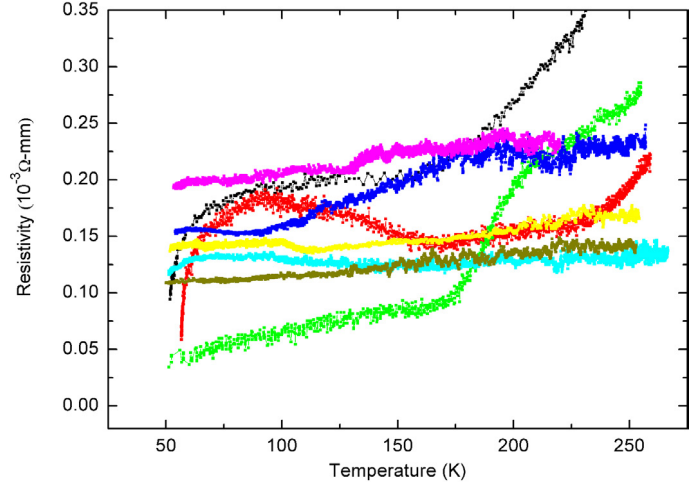

**Figure 3.** Resistivities of PdH$_x$/D$_x$ samples for which no superconducting transition was observed under the same experimental conditions as in Fig. 2.

We propose that the high-temperature hydrogen configuration is not purely octahedral (*oct*), and it has been concluded from neutron diffraction studies[38] that significant tetrahedral (*tet*) D occupancy occurs at 300 °C. In this case it is feasible that some *tet* H/D is preserved at 40 K by rapid cooling. The need for rapid heating to observe the superconducting transition is then easy to understand: diffusion of H/D would drive the system towards the equilibrium configuration, which at room temperature and below is *oct* occupancy[38].

*3.2. Theory*

We now consider theoretically whether *tet* occupancy could lead to enhanced superconductivity by comparing the relevant properties calculated for stoichiometric PdH$^{oct}$ and PdH$^{tet}$. Fig. 2 (a) shows the dependence of the electronic energies of these two systems at zero temperature on the lattice parameter, excluding the zero-point energy. The important conclusion from these calculations is that if the lattice can be expanded sufficiently, then PdH$^{tet}$ is more stable than PdH$^{oct}$. Adding the zero-point energy reverses the order of absolute stability because the potential of the smaller *tet* site is stiffer. However, the energy differences are small enough that some *tet* occupancy is likely at high (or even moderate) temperatures owing to thermal expansion of the lattice and thermal statistics. Moreover, this metastable *tet* occupancy could be maintained upon cooling due to H atoms being trapped by the transition (to *oct*) barrier of 0.1–0.2 eV.

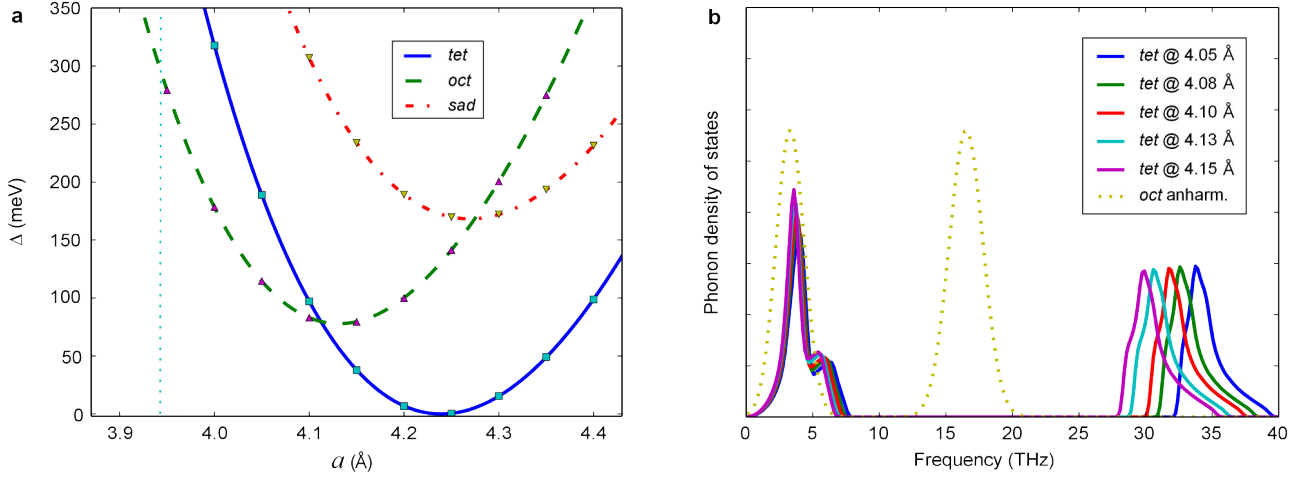

**Figure 4.** *Ab initio* results for energetics and phonons. **(a)** Dependence of electronic energies of PdH$^{tet}$ and PdH$^{oct}$ at 0 K on the lattice parameter. The *bar* index refers to the saddle point on the cube diagonal and provides a simplified measure of the barrier energy. Lattice parameters and relative energies at the minima: $a_{bar}$ = 4.27 Å, $\Delta_{bar}$ = 168 meV; $a_{oct}$ = 4.13 Å, $\Delta_{oct}$ = 78 meV; $a_{tet}$ = 4.24 Å. **(b)** Phonon densities of states (pDOS) for PdH$^{tet}$ sites at different lattice constants. We include representative Gaussians for the pDOS of PdH$^{oct}$ sites taking into account *anharmonic* contributions (at $a \approx 4.1$ Å), taken from [27].

The second important factor is whether the phonon spectrum of PdH$^{tet}$ differs from that of PdH$^{oct}$ so as to enhance the superconductivity. This can happen by increasing the electron–phonon coupling constant ($\lambda$), or by increasing the logarithmically averaged phonon frequency ($\omega_{log}$) (Eqn 3), or some combination of these, since $\lambda$ depends inversely on the phonon frequency (Eqn 2). Recalling that $\alpha(\omega)$ in the Eliashberg spectral function is small at low values of $\omega$, the contribution to $\lambda$ of low-frequency phonon modes (Eqn 2), while scaled by $1/\omega$, is also reduced by $\alpha(\omega)$ approaching zero for small $\omega$, and so medium-frequency modes are the most effective in increasing $\lambda$.

To this end, Fig. 2(b) shows first-principles calculations of the phonon density of states for harmonic PdH$^{tet}$ from this work, compared to anharmonic PdH$^{oct}$ from [27]. Tetrahedral occupancy introduces very-high frequency optical modes that increase $\omega_{log}$ but contribute very little to $\lambda$ (Eqn 2). These are largely insensitive to the lattice parameter, whereas anharmonic octahedral phonons may see substantial lattice effects. On the other hand, *tet* occupancy leads to increased acoustic-mode frequencies in the "Goldilocks zone" of middle frequencies that do contribute substantially to $\lambda$. Thus, even without a detailed Allen-Dynes analysis, it is clear that relocating H/D from *oct* to *tet* sites can readily change the phonon-mediated critical temperature via the pDOS.

The comparison in Fig. 2(b) between the pDOS of *harmonic* PdH$^{tet}$ and *anharmonic* PdH$^{oct}$ is justified by inelastic neutron scattering measurements, summarised in [41]. These showed that for H in the *oct* site of Pd, the potential is strongly well-like, with a positive anharmonicity parameter, whereas for FCC metal-hydrides with *tet* occcupancy, the potential is trumpet-like, with a negative anharmonicity parameter that is less significant compared to the ground-state energy [41]. As shown by Errea et al.[27], anharmonicity of the potential at the *oct* site increases $\omega_{log}$ somewhat but strongly decreases $\lambda$, with the result that $T_c$ is suppressed by a factor approaching ten for H. While the harmonic frequencies for the *tet* site are higher than for the *oct* site because of its smaller size, they are reduced rather than increased by anharmonicity, with an expected much smaller deleterious effect on $\lambda$; in other words, $\lambda$ can plausibly be higher for *tet* occupancy.

## 4. Summary

The accumulated evidence from experiments and theory suggests that the increase in $T_c$ by a factor five owing to our method for preparing PdH$_x$/D$_x$ with $x$ approaching 1 arises in partial occupation of tetrahedral interstitial sites by H/D. It is important that further experiments be performed to first of all confirm this, and then explore ways to make the superconducting state more stable. The likely existence of an isotope effect, together with the metallic character of the material above $T_c$, suggests that this new superconductor is conventional, i.e. phonon mediated. The attainment of such a high $T_c$ for a conventional superconductor without mechanical compression in an interstitial hydride suggests that the search for novel metal-hydride superconductors should be renewed while seeking to increase the density of medium-frequency phonon modes. Further ab initio studies should also be pursued to dig deeper into the fundamental mechanisms of the superconductivity.

**Acknowledgements**

H.M.S. acknowledges receipt of a postgraduate scholarship from Griffith University. The experimental work was supported by a grant from the Australian Research Council.